\documentstyle[11pt,newpasp,twoside,epsf]{article}
\def\edcomment#1{\iffalse\marginpar{\raggedright\sl#1\/}\else\relax\fi}
\reversemarginpar

\begin{document}
\title{Microlensing by free-floating brown dwarfs}
 \author{Hans Zinnecker}
\affil{Astrophysikalisches Institut Potsdam, An der Sternwarte 16,\\
14482 Potsdam, Germany}

\begin{abstract}

We propose a near-infrared microlensing survey of
the central 2 degree field of the Galactic Center,
in an attempt to estimate the surface density and
mass distribution of distant free-floating brown dwarfs
in the bulge and in the disk, acting as lenses of
bright stars towards the Galactic Center. We estimate
the probability (optical depth) of microlensing events
to be §$10^{-7}§$ and the typical timescale (full-width)
of the amplification lightcurve to be about 1 week.
The necessary wide-field NIR survey technology should soon be
available on UKIRT, CFHT, and with VISTA at ESO/Paranal.

\end{abstract}

\section{What is microlensing?}

Microlensing is gravitational lensing without resolving the
split images. Stellar mass lenses are not nearly massive
enough to generate split images, only a "lightcurve" of the
microlensed stars due to the light amplification
associated with the proper motion of the lens can be observed.
The lightcurve is achromatic,
i.e. the same for each wavelength, and also symmetric around
its maximum, two characteristics
which distinguish the microlensing effect from other light
variations (e.g. RR Lyrae stars, Cepheids, or eclipsing binaries).

Microlensing occurs when the observer, the lens
(in our case a brown dwarf), and the source (in our
case a bright star in the Galactic Center or Bulge region)
all lie nearly in the same line of sight.
Microlensing was originally suggested by Paczynski (1986); see
also the excellent review by Paczynski (1996).

\section{Why is microlensing useful in the brown dwarf business?}

As the equation below will show,
microlensing can essentially give us an idea of the {\it mass density}
of field brown dwarfs (or even free-floating giant planets)
in distant parts of the Galaxy, such as in the Bulge, where
star counts can no longer be performed. Microlensing can thus
constrain the substellar initial mass function in places
where we can no longer derive it from a local luminosity function
(e.g. Mera et al. 1998, Peale 1998, Reid 2001, Zinnecker 2001).

In addition, microlensing can in principle (but not in practice)
also yield
the mass of the lens by measuring the full-width of the lightcurve.
Basically the full-width scales with the square root of the mass
of the lens. This is why brown dwarfs with their very low masses
have interestingly short (1 week)
timescales for the light amplification they generate (by moving
across the straight line of sight between the observer and the
source whose brightness is amplified).
However, microlensing cannot determine the mass of the lens
uniquely, as the width of the lightcurve also depends on other
parameters, namely the precise proper motion of the lens in the
plane of the sky and the exact distance to the lens 
and the source from the
observer (see Paczynski 1998 for breaking this degeneracy).

\section{A crash course in microlensing (including basic equations)}

Three ingredients are needed to obtain a crude understanding
of microlensing: the definition of the Einstein radius
(roughly speaking the cross section), the shape
of the light amplification curve as a function of time
including the full-width of that light curve and its
dependence on the lens mass and other parameters,
and finally the expression for the probability, often
called the optical depth, of microlensing.

\begin{figure}[h]
\plotone{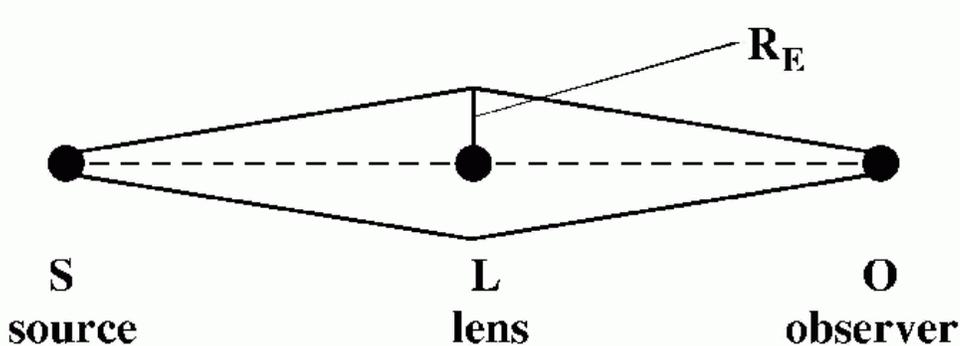}
\caption{Geometry of a gravitational microlensing situation}
\end{figure}

Fig. 1 shows the basic geometry for a microlensing situation
where the lens happens to lie exactly in the S-O line of sight.
$R_E$ is the Einstein radius, defined as

\begin{eqnarray}
R_E&=&\sqrt{\frac{4GM_L}{c^2}D_Sx(1-x)}\nonumber
\end{eqnarray}

\noindent
with $M_L$ the lens mass, $D_s$ the
source distance from the observer, and x the ratio
of lens distance to source distance. ($G$ is
the gravitational constant and $c$ is the speed of light).
Putting numbers into the equation, we find

\begin{eqnarray}
R_E&=&1\,AU\nonumber
\end{eqnarray}

\noindent
for $M_L = 0.03\,M_\odot$, $D_s = 10\,kpc$, $x = 0.5$,
corresponding to a brown dwarf located halfway between
us (the observer) and the Galactic Center, i.e.
a bulge-disk lensing situation. If we want to have
a bulge-bulge lensing situation (''self-lensing``),
we must choose $x = 0.9$ and we obtain $R_E = 0.6\,AU$.
The most relevant $x$ for our purpose lies in between,
as the expression $x(1-x)$ must be weighted with the
mass density distribution of brown dwarfs as a
function of $x$ (see below),
which is higher for larger $x$, as
the density of stars and likely for brown dwarfs
increases towards the Galactic Center.

Next we have to define the impact parameter u (in units of $R_E$)

\begin{eqnarray}
u&=&u(t)=\sqrt{u_{min}^2+\left[\frac{(t-t_0)}{2t_E}\right]^2}\nonumber
\end{eqnarray}

\noindent
where $u_{min}=u(t_0)$ is the distance of closest approach 
(in units of $R_E$)
at time $t_0$ of the moving lens to the line 
of sight between source and observer.
For a closest approach
equal to the Einstein radius ($u_{min}=1$), the source
amplication factor $A$, given by (Paczynski 1986)

\begin{eqnarray}
A&=&\frac{1}{u}\,\frac{u^2+2}{\sqrt{u^2+4}}\nonumber
\end{eqnarray}

\noindent
amounts to $A = 1.34$. For a binary lens (i.e. a brown dwarf or
a planet as companions to a low mass star lens),
the impact parameter must be smaller than $u=1$,
in order to get a bigger amplification and a better
caustic crossing ($u = 0.1$ yields $A \approx 10$).
This is needed to detect a deviation from a single
lens lightcurve; see Mao \& Paczynski (1991), 
Gould \& Loeb (1992), Wambsganss (1997), and 
also Gaudi (these Proc.).
Of course, such
events are correspondingly rarer than $u=1$ events.

Next, we define the duration ($\approx$ 
full-width at half maximum of $A=A(t)$)
of the amplification light curve caused by the lens
proper motion (relative tangential velocity). 
This is given by (Paczynski 1986)

\begin{eqnarray}
t_E&=&\frac{R_E}{v}\nonumber
\end{eqnarray}

\noindent
where $v$ is the proper motion of the lens across the
line of sight towards the source. For $R_E = 0.6\,AU$
(see above) and $v = \sqrt{2}\cdot100\,km/s$ 
(appropriate, given the
Bulge velocity dispersion $\sim100\,km/s$,
similar for both sources and lenses),
we find $t_E \approx 7$\,days.

\section{The probability of microlensing events}

By measuring
the distribution of the duration of
a number of microlensing events one can estimate
the probability for microlensing by brown dwarfs
in the inner Galaxy at any given time. Given a
surface number density of brown dwarfs towards
the Galactic Center, the probability in question
is basically given by the area coverage factor
of all Einstein rings of the field brown dwarfs
taken together, i.e.

\begin{eqnarray}
P&=&\pi R_{E,BD}^2N_{BD}\nonumber
\end{eqnarray}

\noindent
where $N_{BD}$ is the total brown dwarf surface number density.
Inserting the definition of $R_E$ and $x$, we can
rewrite $P$ as

\begin{eqnarray}
P&=&\frac{4\pi GD_S^2}{c^2}\int_0^1\rho(x)x(1-x)dx\nonumber\\
&=&\frac{2\pi}{3}\frac{G\,\overline{\rho}_{BD}}{c^2}D_S^2\nonumber
\end{eqnarray}

Assuming $D_s = 8-10\,kpc$ and replacing $\rho(x)$
by an average brown dwarf mass density $\overline{\rho}_{BD} =
0.015-0.025\,M_\odot/pc^3$ (3\,--\,5 times the local value
in the solar neighborhood),
we can evaluate the expression and obtain

\begin{eqnarray}
P&\approx&10^{-7}\nonumber
\end{eqnarray}

Here we made the implicit assumption that the local
brown dwarf {\it number density} is approximately equal
to the number density of M dwarfs in the solar
neighborhood ($0.1 stars/pc^3$, Reid 1999).
We also took a representative brown dwarf mass of $0.05M_\odot$.
The rest is a scaling of the galactic radial density profile.


\section{From OGLE to IGLE: microlensing in the near-IR}

Microlensing in the near-IR K-band was originally proposed
by Gould (1995). Not aware of this pioneering paper, the idea
occurred to us to extend the optical gravitational microlensing
experiment (OGLE)
from Baade's Window ($l=\pm 1$, $b=-4$ in galactic coordinates) to
a K-band infrared microlensing survey (IGLE) of
the central $2^{\circ}\times\,2^{\circ}$ around the Galactic Centre, a region
where optical studies are hopeless due to the high extinction
($A_v\le30\,mag$); see Catchpole et al. (1990) 
and Philipp et al. (1999) for a  K-band
survey of the area, see also the 2MASS images.

The PI of OGLE is Prof. Udalski of Warsaw University
whose team succeeded
in detecting microlensing events towards the Galactic Bulge,
thanks to a technique described as 'differential image
analysis' (Alard \& Lupton 1998) with which time-variable
sources, magnified by microlensing, could be convincingly
recognized in large and very crowded CCD fields 
(Udalski et al. 2000, Wozniak et al. 2002).
A similar technique should be applied to near-IR imaging
surveys to beat the even higher crowding towards the
central Galactic Bulge region.

\section{Bulge giants lensed by Bulge brown dwarfs}

By concentrating on the inner Galactic Bulge (Galactic
Centre) region we maximise the number density of bright
red giant sources to be lensed by faint brown dwarfs also
in the Bulge (bulge-bulge lensing,
see Kiraga \& Paczynski 1994). Red giants have an absolute
K-magnitude close to -3\,mag, thus for a distance modulus
$m-M=14.5\,mag$ to the Galactic Centre and an extinction of
$A_K=3\,mag$ (corresponding to $A_v=30\,mag$) the apparent
magnitude of red giants near the Galactic Centre becomes
$K=14.5\,mag$. On a 4\,m class telescope, a S/N=20 detection
of a 10\,\% brightness variation in sub-arcsec seeing 
takes about 2\,min, so near-IR surveys with 4\,m class
telescopes can proceed fast enough to observe an area of
$2^{\circ}\times2^{\circ}$ several times a day for several days, enough
to catch microlensing events of short duration (few days);
such a survey will have other useful spin-offs 
(e.g. probing galactic bar structure: see
Peale, 1998; Evans \& Belokurov, 2002).

\section{Telescopes/instruments available}

For the purposes of the proposed near-IR microlensing survey
towards obscured central Galactic Bulge sources, there are
practically 4 choices of telescopes and instrumentation:

\begin{itemize}
\item
the wide-field camera WFCAM for UKIDSS at UKIRT (Warren 2002), ready
for use in 2004, with a 2\,$\times$\,2 arrangement 
of 2\,k\,$\times$\,2\,k detectors
at $0.4"/pix$ (total instantaneous FOV per exposure 
$\approx0.2\,\Box^{\circ}$)\\

\item
the wide-field camera WIRCAM at CFHT (Beuzit, these Proc.),
ready for use in 2004, with a total instantaneous FOV per exposure
of about $0.1\,\Box^{\circ}$ (4\,k\,$\times$\,4\,k\,=\,20'\,$\times$\,20') 
at $0.3"/pix$\\

\item
the planned $1\,\Box^{\circ}$ FOV telescope VISTA at ESO/Paranal, to be ready
in 2006, with a 4\,$\times$\,4 arrangement of 
2\,k\,$\times$\,2\,k near-IR detectors at
$0.34"/pix$ (practically a 2\,$\times$\,2 extension of WFCAM for UKIDSS at UKIRT)\\

\item
the NGST near-infrared camera, planned for 2010, with two 4\,k\,$\times$\,4\,k
detector at $0.04"/pix$ in the K-band, 
i.e. a total FOV of about 3'\,$\times$\,3'.\\
\end{itemize}

The two immediate options (at UKIRT/CFHT) each provide some
16 million pixels per exposure, while the two medium-term
possibilities (VISTA/NGST) feature 64 and 48 million pixels.
In order to cover the $2^{\circ}\times\,2^{\circ}$ 
Galactic Centre area, we  need
about 20, 40, and 4 exposures with WFCAM, WIRCAM, and VISTA,
respectively. In each of these cases\footnote
{NGST is different. It will not observe a large FOV,
but will operate mostly in staring mode and will  go
much deeper. The hope is to resolve Bulge subgiants and 
Main Sequence stars.},
we can sample on the
order of $3\cdot10^8$ pixels and complete the 
high S/N observations of the bright bulge giants
within 1\,hr (integration
plus read-out time). We suggest 3 sets of observations per
night for 14 consecutive nights. This should be good enough
to probe the previously estimated probability of $10^{-7}$
for microlensing by bulge brown dwarfs.

\section*{Acknowledgment}

Travel support from the Deutsche Forschungsgemeinschaft (DFG)
is gratefully acknowledged. I also thank 
Bogdan Paczynski, Penny Sackett, Robert Schmidt and Joachim
Wambsganss for stimulating discussions.

\section*{Note added in proof}

Delplancke et al. (2001) discussed the possibilities to resolve
the images of microlensed objects towards the Galactic Bulge by
the use of long baseline interferometry, in particular the ESO-VLTI
(K-band observations with the AMBER and PRIMA instruments).
The ability to measure the angular separation between the
microlensed images (of the order of 1 milli-arcsec) 
and to measure the brightness of the images 
as a function of time will enable
a direct and unambiguous determination of the lens masses
and locations, as well as their proper motions.
This then allows one to break the degeneracy (mass, tangential
velocity, distance of the lens) inherent in the
interpretation of the width of the lightcurve of a microlensing
event.

\end{document}